\documentclass[prl,twocolumn,showpacs]{revtex4}
\usepackage{graphicx}

\begin{document}

\newcommand {\ee}[1] {\label{#1} \end{equation}}
\newcommand{\be}{\begin{equation}}
\newcommand {\e} {\varepsilon}
\def\reff#1{(\ref{#1})}
% \draft
% \tightenlines

% \wideabs{
\title{System size resonance in coupled noisy systems and in the Ising model}
\date{\today}
\author{A. Pikovsky}
\author{A. Zaikin}
\affiliation{Department of Physics, University of Potsdam,
	Postfach 601553, D-14415 Potsdam, Germany}
\author{ M. A. de la Casa}
\affiliation{Departamento F{\'{\i}}sica Fundamental, 
	Universidad Nacional de Educaci{\'o}n a
Distancia, 28040 Madrid, Spain}

\begin{abstract}%
We consider an ensemble of coupled nonlinear noisy oscillators
demonstrating in the thermodynamic limit an Ising-type transition. In the
ordered phase and for finite ensembles stochastic flips of the mean field
are observed with the rate depending on the ensemble size. When a  small periodic force
acts on the ensemble, the  linear response of the system has a maximum at a
certain system size, similar to the stochastic resonance phenomenon. We demonstrate
this effect of system size resonance for different types of noisy oscillators and
for different ensembles -- lattices with nearest neighbors coupling and globally
coupled populations. The Ising model is also shown to demonstrate the system
size resonance. 
\end{abstract}

% }
%\twocolumn
\maketitle

% Introduction
Stochastic resonance has attracted much interest 
recently~\cite{GJ}. As was
demonstrated in~\cite{Benzi-Sutera-Vulpiani-81}, a response of a noisy nonlinear
system to a periodic forcing can exhibit a resonance-like dependence on the
noise intensity. In other words, there exists a ``resonant'' noise intensity at
which the response to a periodic force is maximally ordered. Stochastic
resonance has been observed in numerous
experiments~\cite{LS}. Noteworthy, the
order in a noise-driven system can have a maximum at a certain noise level even
in the absence of periodic forcing, this phenomenon being called coherence
resonance~\cite{PN}.

Being first discussed in the context of a simple bistable model, stochastic
resonance has been also studied in complex systems consisting of many elementary
bistable
cells~\cite{JM}, 
%Again, one observes a resonance-like dependence on the
%noise intensity
moreover, the resonance may be enhanced due to 
coupling~\cite{Lindner}. In this paper 
we discuss another type of resonance in such
systems, namely the {\em system size resonance}, 
when the dynamics is maximally
ordered at a certain number of interacting subsystems.  Contrary to 
previous reports of array-enhanced stochastic
resonance 
%phenomena 
(cf. also~\cite{NH}), 
here we fix the noise strength, coupling, and other parameters; only the
the size of the ensemble changes.

The basic model to be considered below is the ensemble of noise-driven bistable
overdamped oscillators, governed by the Langevin equations
\be
\dot{x}_i=x_i-x_i^3+\frac{\e}{N}\sum_{j=1}^N
(x_j-x_i)+\sqrt{2D}\xi_i(t)+f(t)\;.
\ee{ens1}
Here $\xi_i(t)$ is a Gaussian white noise with zero mean: $\langle
\xi_i(t)\xi_j(t')\rangle = \delta_{ij}\delta(t-t')$; $\e$ is the
coupling constant; $N$ is the number of elements in the ensemble, and $f(t)$ is
a periodic force to be specified later. In the absence of periodic force 
the model
\reff{ens1} has been extensively studied in the thermodynamic limit $N\to
\infty$. It demonstrates an Ising-type phase transition at $\e=\e_c$ from the disordered state
with vanishing mean field 
$X=N^{-1}\sum_i x_i$
to the ``ferromagnetic'' state
with a nonzero mean field $X=\pm X_0$ 
%A theory of this transition, based on the
%nonlinear Fokker-Planck equation, was developed in 
\cite{Desai-Zwanzig-78}. 
%where also the
%expressions for the critical coupling $\e_c$ are given. 

While in the thermodynamic limit the full description of the dynamics is
possible, for finite system sizes we have mainly a qualitative picture:
in the ordered phase the mean field $X$ 
switches between the values $\pm X_0$ and its average 
vanishes for all couplings. The rate of switchings depends on the
system size and tends to zero as $N\to \infty$.
% The asymptotic dynamics in this
% limit has been discussed in 
% \cite{Dawson-Gaertner-87}.

For us, of the main importance is the fact that qualitatively the behavior of the
mean field can be represented as the noise-induced dynamics 
in a potential with one minimum in the disordered phase (at $X=0$) 
and two
symmetric minima (at $X=\pm X_0$) in the ordered phase.
Now applying the ideas
of the stochastic resonance, one can expect in the
bistable case (i.e. in the ordered phase for small enough noise or for large
enough coupling) a resonant-like
behavior of the response to a periodic external force 
when the intensity of the effective noise is changed. Because this
intensity is inverse proportional to $N$, we obtain the resonance-like curve of the
response in dependence of the system size.
The main idea behind the system size resonance is that in finite ensembles of
noise-driven or chaotic systems the dynamics of the mean field can be represented
as driven by the effective noise whose variance is inverse proportional to the
system size~\cite{DPH}.
This idea has been applied to description of a transition to collective
behavior in ~\cite{Pikovsky-Ruffo-99}. 
In ~\cite{Pikovsky-Rateitschak-Kurths-94}
it was demonstrated that the finite-size fluctuations can cause a transition
that disappears in the thermodynamic limit. 
%The description of finite-size
%effects in deterministic chaotic systems using the effective noise concept 
%has been
%suggested in~\cite{}. 
%We emphasize that noise plays  an essential role in this picture: with $D=0$
%\reff{ens1}
%is a deterministic oscillator (double or single well, depending on $\e$), 
%whose response to a periodic force does not depend on $N$.

Before proceeding to a quantitative analytic description of the phenomenon, we illustrate
it with direct numerical simulations of the model \reff{ens1}, with a
%sinusoidal 
forcing term $f(t)=A\cos(\Omega t)$. Figure \ref{fig:ens1} shows the
 linear response function, i.e. the ratio of the spectral component in the mean
field at frequency $\Omega$ and the amplitude of forcing $A$, in the limit 
$A\to 0$. For a given frequency  $\Omega$ the dependence on the system size is a
bell-shaped curve, with a pronounced maximum. 
 The dynamics of the mean field $X(t)$ is
illustrated in Fig.~\ref{fig:ens2}, for three different system sizes. 
%and for a
%particular frequency. 
The
resonant dynamics (Fig.~\ref{fig:ens2}b)
demonstrates a typical for stochastic resonance synchrony between the driving
periodic force and the switchings of the field between the two stable positions.
%For non-resonant conditions 
%(Fig.~\ref{fig:ens2}a,c) the switchings are either too frequent or too rare, as
%a result the response is small.

\begin{figure}[!htb]
  \centerline{\includegraphics[width=0.45\textwidth]%
  {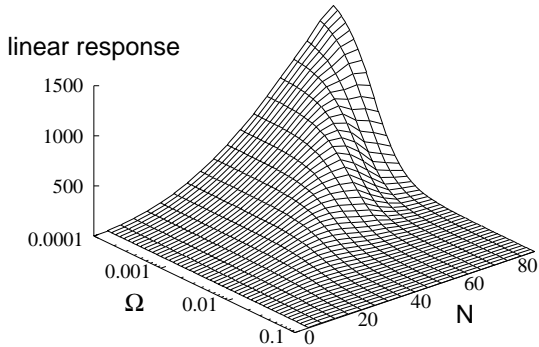}}
  \caption{Linear response of the ensemble \reff{ens1}($D=0.5$, $\e=2$) 
  in dependence on the frequency  and the system
  size $N$. 
  %The response to forces with smaller frequencies 
  %is shifted to larger
  %system sizes, where the effective noise, and, consequently, the switching
  %rate, is smaller. The linear response is obtained by virtue of the
  %fluctuation-dissipation theorem.
  }
  \label{fig:ens1}
\end{figure}

\begin{figure}[!htb]
  \centerline{\includegraphics[width=0.45\textwidth]%
  {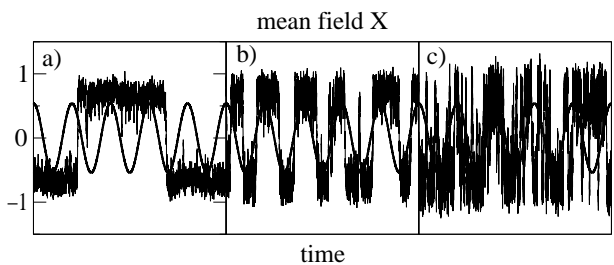}}
  \caption{The time dependence of the mean field in the ensemble \reff{ens1} for
  $D=0.5$, $\e=2$, $A=0.02$, $\Omega=\pi/300$, and different sizes of the ensemble: (a) $N=80$,
  (b) $N=35$, and (c) $N=15$. We also depict the periodic force (its
  amplitude is not in
  scale) to demonstrate the synchrony of the switchings with the forcing in (b).}
  \label{fig:ens2}
\end{figure}

To describe the system size resonance analytically, we use, 
following~\cite{Desai-Zwanzig-78}, 
the Gaussian approximation. In this approximation one writes $x_i=X+\delta_i$
and assumes that $\delta_i$ are independent Gaussian random variables with zero
mean and the variance $M$. Assuming furthermore that $N^{-1}\sum_i
\delta_i^2=M$
and neglecting the odd moments $N^{-1}\sum_i \delta_i$, $N^{-1}\sum_i
\delta_i^3$, as well as the correlations between $\delta_i$ and
$\delta_j$, we obtain from \reff{ens1} the equations for $X$ and $M$:
\begin{eqnarray}
\dot{X}&=&X-X^3-3MX+\sqrt{\frac{2D}{N}}\eta(t)+f(t)\;,\label{ga1}\\
\frac{1}{2}\dot{M}&=&M-3X^2M-3M^2-\e M+D\;,\label{ga2}
\end{eqnarray}
where $\eta$ is the Gaussian white noise having the same properties as
$\xi_i(t)$.
In the thermodynamic limit $N\to\infty$ the noisy term $\eta$ vanishes. If the
forcing term is absent ($f=0$),  the
equations coincide with those derived in~\cite{Desai-Zwanzig-78}. 
This system of coupled
nonlinear equations 
exhibits a pitchfork bifurcation of the equilibrium $X=0,\;M>0$ at $\e_c=3D$. 
This bifurcation is supercritical for $D>2/3$ in accordance with the exact
solution of \reff{ens1} given in \cite{Desai-Zwanzig-78}, 
below we consider only this case.
For
$\e>\e_c$ the system is bistable with two symmetric stable fixed
points 
\be
X_0^2= (2-\e+S)/4\;,\quad
M_0=(2+\e-S)/12\;,
\ee{xnul}
(here $S=\sqrt{(2+\e)^2-24D}$) 
and the unstable point $X=0,\;M=(1-\e+\sqrt{(1-\e)^2+12D})/6$.
Now, with the external noise $\eta$ and with the periodic force $f(t)$ the
problem reduces to a standard problem in the theory of stochastic resonance,
i.e. to the problem 
of the response of a noise-driven  nonlinear bistable system to an external
periodic force (because the noise affects only the variable $X$, it does not
lead to unphysical negative values of variance $M$, since $\dot{M}$ is strictly
positive at $M=0$). 
%This response has a maximum at a certain noise intensity, which
%according to \reff{ga1} is directly related to the system size. 

 To obtain an analytical formula, we perform further simplification of the
system \reff{ga1},\reff{ga2}. Near the bifurcation point 
we can use the slaving principle to
%the dynamics of $X$ is slower
%than that of $M$, and we can exclude the latter one assuming $\dot{M}\approx 0$.
%Then from \reff{ga2} we can express $M$ as a function of $X$ and substitute to
%\reff{ga1}. Near the bifurcation point we 
obtain a standard noise-driven bistable system
\be
\dot{X}=aX-bX^3+\sqrt{\frac{2D}{N}}\eta(t)+f(t)\;,
\ee{gast}
where $a=1+0.5(\e-1)-0.5\sqrt{(\e-1)^2+12D}$,
$b=-0.5+1.5(\e-1)((\e-1)^2+12D)^{-1/2}$. A better approximation valid also
beyond a vicinity of the critical point can be constructed if we use
$\bar{b}=aX_0^{-2}$ instead of $b$, where the fixed point $X_0$ is given by
\reff{xnul}. Having written the ensemble dynamics as a standard noise-driven
double-well system~\reff{gast}
(cf.~\cite{GJ,Jung-Hanggi-91}), we can use the
analytic formula for the linear response $R$ derived in ~\cite{Jung-Hanggi-91}. It
reads
\be
R=\frac{N X_0^2}{2Da}\left(\frac{{\cal D}_{-3/2}(-\sqrt{s})}{{\cal
D}_{-1/2}(-\sqrt{s})}\right)^2
\left[1+\frac{\pi^2\Omega^2}{2 a^2}\exp(s)\right]^{-1}
\ee{lrf}
where $s=aNX_0^2/(2D)$, and ${\cal D}$ are the parabolic cylinder functions.
We compare the theoretical linear response function with the numerically
obtained one in Fig.~\ref{fig:ga}. The qualitative correspondence is good,
moreover, the maxima of the curves are rather good reproduced with the formula
\reff{lrf}. 
%This shows that the resonant system size is quite good
%quantitatively described by
%the Gaussian approximation, see Fig.~\ref{fig:ga}.

Above we concentrated on the properties of the linear response. Numerical
simulations with the finite forcing amplitude yielded the results similar to
that presented in Figs.~\ref{fig:ens1},\ref{fig:ga}. However, for large
amplitudes of forcing (e.g., $A>0.1$ for $\Omega=0.01$, $D=0.5$, $\e=2$) 
a saturation was observed: here 
the response grows monotonically with $N$. This is in full agreement with the
corresponding property of the stochastic resonance in double-well systems of
type \reff{gast}, where the saturation occurs for small noise intensities (cf.
Fig. 7 in \cite{GJ}), 
%Qualitatively, the saturation is 
due to the disappearance of
multistability for large forcing amplitudes. 
%so that the oscillator \reff{gast}
%switches at every period of the forcing, contrary to the case of small
%amplitudes, where the switchings are rare (Fig.~\ref{fig:ens2}a). 
%The critical
%value of the amplitude, at which the bistability disappears, decreases
%monotonically with the driving frequency and is $\approx 0.37$ at $\Omega\to 0$
%(to be compared with the values $1.1$ and $0.4$ for the frequencies $1$ and
%$0.1$, respectively). 

\begin{figure}[!htb]
  \centerline{\includegraphics[width=0.45\textwidth]%
  {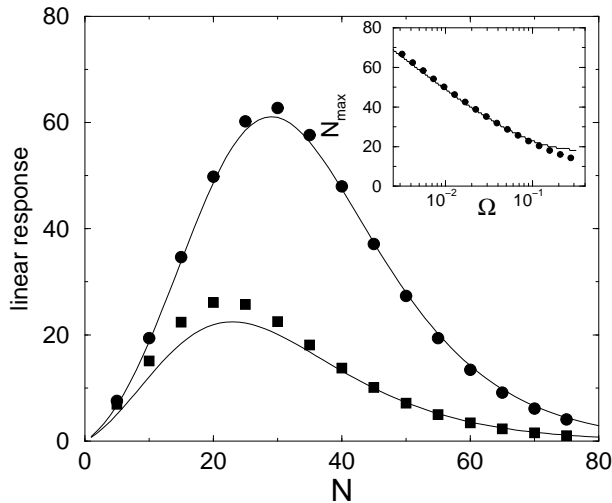}}
  \caption{Comparison of the system-size dependencies of the linear response
  function for frequencies $\Omega=0.05$ (circles) and $\Omega=0.1$ (squares)
  with theory \reff{lrf}. The parameters are $D=1$ and $\e-\e_c=2.5$
  (where the the exact $\e_c$ and the approximate $\e_c=3D$ are used for the
  ensemble and the Gaussian approximation, respectively). Inlet: 
  Dependence of the system size yielding maximal linear response on the
  driving frequency $\Omega$ (circles: simulations of the ensemble \reff{ens1},
  line is obtained by maximizing the expression \reff{lrf}).}
  \label{fig:ga}
\end{figure}

It is instructive to compare the response of the noise-driven
system~\reff{ens1} with the noise-free case $D=0$. Without external force, the
ensemble relaxes eventually to a steady state solution with some mean field $X$;
in this state each oscillator can be in one of stable steady positions of the
potential, 
%given by the
%roots of $x(1-\e)-x^3+\e X=0$. For $\e< 1$ this equation can have two stable
%roots ($y_1$ and $y_3$), 
correspondingly the oscillators form one or two
clusters.
%: a portion $p_1$ is in the state $y_1$ and a portion $p_3=1-p_1$ is
%in the state $y_3$. The partition is determined self-consistently from the
%relation $X=p_1y_1+p_3y_3$. For $\e>1$ there is only one stable root,
%thus all oscillators are in one cluster and $X=\pm 1$. 
From the clustering it
follows that the linear response does not depend on the number of elements in
the ensemble. 
%Indeed, in both cases of one or two clusters, the system
%\reff{ens1} can be reduced to a system of one or two equations, because the
%dynamics of all elements in a cluster is identical. The linear response of a
%steady state in these equations can be readly found (e.g., it is
%$R=(4+\Omega^2)^{-1}$ for the case of one cluster).
%, and
%\begin{widetext}
%\[
%R=\left|\frac{i\Omega-1+3(y_1^2p_3+y_3^2p_1)+\e}
%{-\Omega^2-i\Omega[2-3(y_1^2+y_3^2)-\e]+(1-3y_1^2+\e(p_1-1))
%(1-3y_3^2+\e(p_3-1))-\e^2p_1p_3}\right|^2 
%\]
%\end{widetext}
%for the case of two clusters. 
Our numerical experiments demonstrated also that
the response is system-size independent for large forcing amplitudes as well,
where, e.g., the force-induced cluster-mergings occur. Thus, the effect of
system size resonance essentially relies on the presence of noise, which breaks
the clustering.
%In particular, one can say that for a certain system size a maximally correlated
%response of noncoherently driven oscillators is achieved.

Above we have considered the system of globally coupled nonlinear 
oscillators~\reff{ens1}.
The same effect of system size resonance can be observed in a lattice with
nearest neighbors coupling as well. 
%Then, instead of \reff{ens1}, we have
%\be
%\dot{x}_i=x_i-x_i^3+\frac{\e}{K}\sum_{\langle ij\rangle} 
%(x_j-x_i)+\sqrt{2D}\xi_i(t)+f(t)\;,
%\ee{lat1}
%where the number of nearest neighbors $K$ depends on the geometry of the
%lattice and on the dimension of the space. 
In the thermodynamic limit, 
the Ising-type phase transition occurs  in the
lattice (if its dimension is larger than one). 
Similar to the globally coupled
ensemble, in finite lattices in the ordered phase the switchings between the two
stable states of the mean field are observed. With the same argumentation as
above we can
conclude that the response of the mean field to a periodic forcing 
%$f(t)$ 
can
have a maximum at a certain lattice size, while all other parameters (noise
intensity, coupling strength, etc.) are kept constant. We illustrate this in
Fig.~\ref{fig:lat}. 
%Here the two-dimensional lattice with periodic
%boundary conditions is studied. In order to keep the lattice of nearly 
%quadratic form, we have chosen the lattices with sizes either $l^2$ or 
%$l(l+1)$, with
%$l=2,3,4,\ldots$. The response to the periodic force has a pronounced maximum at a
%certain size of the system.

\begin{figure}[!htb]
  \centerline{\includegraphics[width=0.45\textwidth]%
  {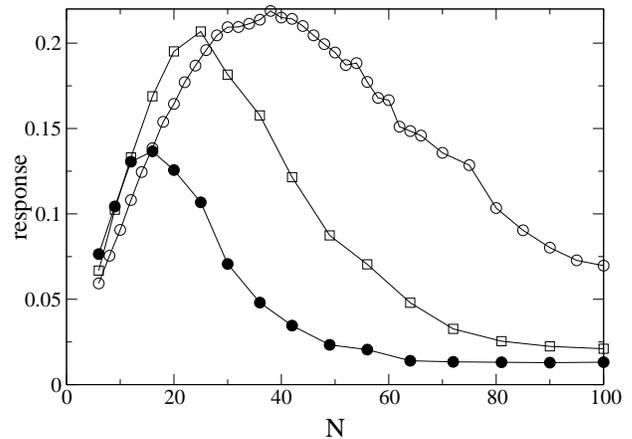}}
  \caption{Filled circles: Response of a  two-dimensional lattice of $N$ 
  with nearest-neighbors coupling  for
  $A=0.02$, $T=500$, $D=0.5$, and $\e=4$. Squares: 
  Response of system \reff{ens2} 
   (a two-dimensional lattice with $D=1.25$, $\e=30$,
  $A=0.1$ and  $T=140$). Circles: the same as squares, but for
  a globally coupled lattice 
   with  $D=1$, $\e=20$, $A=0.1$ and  $T=100$.}
  \label{fig:lat}
\end{figure}

 As the next example we consider the two-dimensional nearest neighbor Ising model
in the presence of a time-dependent external field. The Hamiltonian of the
system reads
\be
H=-J\sum_{\langle ij\rangle}s_is_j-A\cos(\Omega t)\sum_i s_i\;,
\ee{is1}
where $J>0$ and $s_i=\pm 1$. We are interested in the dependence of the 
response of  the mean magnetization
$m(t)=\frac{1}{N}\sum_i s_i(t)$
on the system size $N$ (for the usual stochastic resonance in the Ising model,
i.e., for the dependence of the response on the temperature, 
see \cite{NP}). To calculate the linear response we used the
fluctuation-dissipation theorem and obtained this quantity by virtue of the
power spectrum of fluctuations of $m(t)$. The latter was found using the
Metropolis Monte Carlo method on a lattice with helical boundary
conditions~\cite{Newman-Barkema-99}. The results presented in
Fig.~\ref{fig:ising} demonstrate the system size resonance of the linear
response in the two-dimensional Ising model.

\begin{figure}[!htb]
  \centerline{\includegraphics[width=0.45\textwidth]%
  {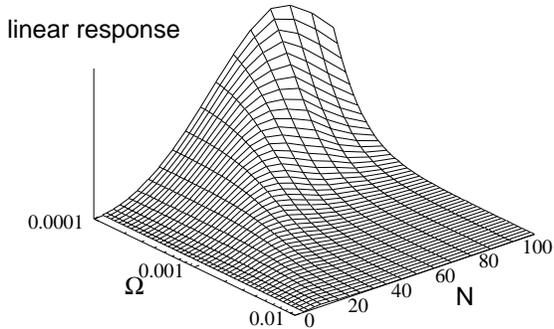}}
  \caption{Linear response (in arbitrary units) 
  of the Ising model \reff{is1} for the temperature $T=2J$  slightly below the
  critical temperature $T_c=2.269 J$.} 
  \label{fig:ising}
\end{figure}

As the last example of the system size resonance we consider a lattice where
each individual element does not exhibit bistable noisy dynamics, but such a
behavior appears due to interaction and multiplicative noise. 
This model is described by the set of
Langevin
equations~\cite{VB,Landa-Zaikin-Schimansky-Geier-98}
\be
\dot{x}_i=-x_i(1+x_i^2)^2+\frac{\e}{K}\sum_{j}
(x_j-x_i)+\sqrt{2D}\xi_i(t)(1+x_i^2)+f(t)\;.
%A\cos(2\pi t/T)\;.
%\label
\ee{ens2}
%The difference to the model \reff{ens1} is that the noise is multiplicative and
%the on-site potential has only one minimum. $K$ is the number of elements to
%which the oscillator $i$ is coupled, for global coupling $K=N$ and for a lattice
%of dimension $d$ with nearest-neighbors coupling $K=2d$. 
As has been demonstrated 
in~\cite{VB},
in some region of couplings $\e$ system \reff{ens2} exhibits the Ising-type
transition.
% characterized in the thermodynamic limit $N\to\infty$  by the onset
%of nonzero mean field $X$. Due to the symmetry of \reff{ens2},
%there are states with positive and negative mean field. 
If an additional
additive noise is added to \reff{ens2}, then one observes transitions between
these states and the so-called double 
stochastic resonance in the presence of the periodic
forcing~\cite{ZK}. As is evident from the considerations above, such transitions
occur even in the absence of the additive noise if the system is finite. Thus,
the system size resonance should be observed in the lattice   \reff{ens2} as
well. We confirm this in Fig.~\ref{fig:lat}. 
%Note that 
%in ~\cite{Zaikin-Murali-Kurths-01} a realistic electronic circuit modeling the
%ensemble similar to \reff{ens2} 
%is described, providing a possible experimental realization
%of the effect.

Another possible field of application of the system size resonance is the neuronal
dynamics (see, e.g., \cite{Tass-99}). 
Individual neurons have been demonstrated to exhibit stochastic
resonance~\cite{LS,DR}.
While in experiments one can easily adjust noise to achieve the
maximal sensitivity to an external signal, it may be not obvious how this
adjustment takes place in nature. The above consideration shows, that changing
the number of elements in
a small ensemble of coupled bistable elements to the optimum 
can significantly improve the
sensitivity (cf.~\cite{JM}). Moreover, changing its 
connectivity and/or coupling strength, a
neuronal system can tune itself to signals with different frequencies. 

Concluding, we have shown that in populations of coupled 
noise-driven elements, exhibiting in the thermodynamic limit the Ising-type
transition, in the ordered phase (i.e. for relatively small noise and large
coupling)
the response to a periodic force achieves maximum at a
certain size of the system. We demonstrated this effect for the Ising model, as
well as for lattices and globally coupled ensembles noisy oscillators.
We expect the system size resonance to occur also in purely 
deterministic systems demonstrating the Ising-type transition, e.g. in the
Miller-Huse coupled map lattice~\cite{Miller-Huse-93}.
 The system size resonance 
is described theoretically by reducing the dynamics of
the mean field to a low-dimensional bistable model with 
an effective noise that is
inverse proportional to the system size. The stochastic resonance in the mean
field dynamics then manifests itself as the system size resonance.

We thank N. Brilliantov, A. Neiman, J. Parrondo, F. J. de la Rubia and R. Toral 
for useful discussions and M. Rosenblum for help in analysis of eq.~\reff{lrf}. 
M. C. thanks the DGESEIC Project No. PB97-0076 and the
DGES Grantship No. AP98 07249358. A.Z. acknowledges financial support from ESA.

\end{document}